# From the time series to the complex networks: the parametric (dynamical) natural visibility graph

I.V. Bezsudnov[1], A.A. Snarskii[2,3]

[1]Nauka – Service JSC, Moscow, Russia

[2]Dep. of Gen. and Theor. Physics, National Technical University of Ukraine "KPI", Kiev, Ukraine

[3]Institute for Information Recording NAS Ukraine, Kiev, Ukraine

Abstract

We present the modification of natural visibility graph (NVG) algorithm used for the mapping of the time series to the complex networks (graphs). We propose the parametric natural visibility graph (PNVG) algorithm. The PNVG consists of NVG links, which satisfy an additional constraint determined by a newly introduced continuous parameter - the view angle. The alteration of view angle modifies the PNVG and its properties such as the average node degree, average link length of the graph as well as cluster quantity of built graph etc. We calculated and analyzed different PNVG properties depending on the view angle for different types of the time series such as the random (uncorrelated, correlated and fractal) and cardiac rhythm time series for healthy and ill patients. Investigation of different PNVG properties shows that the view angle gives a new approach to characterize the structure of the time series that are invisible in the conventional version of the algorithm. It is also shown that the PNVG approach allows to distinguish, identify and describe in detail various time series.

---------------------------

Corresponding author: Dr. Igor Bezsudnov, e-mail: biv@akuan.ru

# I. INTRODUCTION

The idea to investigate the time series by mapping them to the complex networks (graphs) is very attractive. Two advanced research areas are combined under this approach: the methods of the nonlinear time series analysis [1-8] and the theory of the complex networks [9-15]. There is an opportunity to apply the rich, well-developed methods of the complex networks analysis to the investigation of the time series with a complicated structure, such as the fractal time series.

Currently there are several algorithms for mapping the time series to complex networks. For instance, it was suggested in [16] to build a network using the proximity of the coordinates in the Poincare section of the time series (see also [17-20]). Visibility mapping algorithms were proposed by *L.Lacasa et al* in [21,22] for constructing a Natural Visibility Graph (NVG) [21] and Horizontal Visibility Graph (HVG) [22].

Let $\{x(t_i), i=1..N\}$ be a time series of $N$ data, $t_i$ are in natural temporal ordering. The NVG [21] is created by mapping of a time series of $N$ data to a network (graph) of the $N$ nodes. The link $(i, j)$ belongs NVG if on the time series plot $x(t_k)$ for all $t_k$ between $t_i$ and $t_j$ are below the line connecting $x(t_i)$ and $x(t_j)$ (see Fig.1*a,b*). The HVG algorithm [22] is similar to the NVG algorithm but has a modified mapping criterion. The link $(i, j)$ exists in the HVG only if all $x(t_k)$ for $t_k$ between $t_i$ and $t_j$ are less than both $x(t_i)$ and $x(t_j)$ (see Fig.1*c, d*).

NVG and HVG are the connected graphs, and the HVG is a subgraph of the NVG. The HVG and the NVG are invariant under affine transformations of time series plot [21, 22].

The use of the NVG and the HVG algorithms allows us to describe and explore the time series of complex structure associated with a variety of phenomena: fluctuations of turbulent flows [23], stock market indices [24], human heartbeat dynamics [25, 26], stochastic and chaotic series [27 - 30] and others [31 - 37].

In this paper we introduce the parameter $\alpha$ which we call "View angle" and describe the algorithm to create the Parametric Natural Visibility Graph (PNVG) whose mapping criteria depends on view angle $\alpha$. The PNVG algorithm consists of three major steps: first we build NVG [21], then we set the natural temporal direction to each NVG link and set the its weight equal to the angle between link and downward directions (see Fig.1a) and finally we construct PNVG from NVG links which has weight (angle) less than the given view angle $\alpha$. The PNVG is a directed subgraph of the underlying NVG but PNVG is not necessarily connected graph.

Actually, the ability to change arbitrarily view angle $\alpha$ adds the word "parametric" to the name of the algorithm. We will also use the abbreviation PNVG($\alpha$). The PNVG algorithm creates a new graph for each view angle $\alpha$. In such a way we are able to investigate the properties of the PNVG($\alpha$) how they depend upon view angle $\alpha$.

In the article, the mapping criterion for the Parametric Natural Visibility Graph algorithm is proposed and relation between PNVG, NVG and HVG is discussed. Then we introduce and investigate the behavior of PNVG($\alpha$) properties: the average nodes degree, the average link length of the graph and the number of clusters in the PNVG($\alpha$). Different artificial random time series are considered i.e. with and without correlations and, more complicated, having fractal dimension too. At the end, we show an example of application of the PNVG algorithm to the heart beat RR intervals analysis and show the possibility to classify different types of heart failure.

## II. MODEL AND ALGORITHM

We start the construction of the PNVG from a sequential time series $\{x(t_i), i=1..N\}$ of $N$ data such as lighting event times or solar flare or earthquake magnitudes etc.

The PNVG mapping algorithm is as follows:

1. To build NVG [21] as it was described above using common NVG criteria for mapping

$$\left\{(i,j) \in NVG, \quad x(t_k) < x(t_i) + \left(x(t_j) - x(t_i)\right)\frac{t_k - t_i}{t_j - t_i}, \quad i < k < j\right\}, \qquad (1)$$

where $i$ and $j$ are numbers of two arbitrary time events $t_i < t_j$ and $t_k$ any event between $t_i < t_k < t_j$.

2. To set for every link of NVG the direction and weight (angle).

All the NVG links have a natural temporal direction, i.e. the link $(i, j)$, $i < j$ is considered to be directed from $i$ to $j$. The weight value is the angle on the time series plot between the downward direction and the direction of the line going from $x(t_i)$ to $x(t_j)$ (see Fig.1a)

$$\alpha_{ij} = arctg \frac{x(t_j) - x(t_i)}{t_j - t_i}, \quad i < j. \tag{2}$$

3. To select links from created directed and weighted graph according to the rule that uses introduced parameter – view angle $\alpha$

$$\{(i, j) \in \text{PNVG}(\alpha), \ \alpha_{ij} < \alpha\}. \tag{3}$$

One can find from above rules that PNVG($\alpha$) is a directed acyclic graph, PNVG($\alpha$) is subgraph of underlying NVG, therefore PNVG($\alpha$) is invariant under the affine transformations of time series plot and PNVG($\alpha$) can be either connected or disconnected graph.

Authors have to emphasize that results of rule 2 depend on the used scale for time series plot $x(t_i)$ i.e. if scale is altered then the angles $\alpha_{ij}$ are changed and therefore rule 3 selects different links. This uncertainty is eliminated if we use some kind of natural initial scale and appropriate view angle transformations. In this paper we deal only with time series that have same scales in both directions, i.e. $x(t_i)$ and $t_i$ have the same units and the same scales both in horizontal and in vertical directions of time series plot, so $\alpha_{ij}$ has fixed value in any scale.

Such equally scaled time series can be built for instance from sequential time intervals set $\{d_i, i=1..N\}$ like RR heart beat intervals. Using initial $d_i$ data we restore event times as $\{t_1 = 0, t_i = t_{i-1} + d_{i-1}, i=1..N+1\}$ and create time series $\{x(t_i) = t_{i+1} - t_i, i=1..N\}$, in such a case $x(t_i)$ values are always positive and event $x(t_i)$ occurs at time $t_i$.

Let us look to the mapping illustration presented in Fig.1 for fragment of the time series $x(t_i)...x(t_{i+8})$

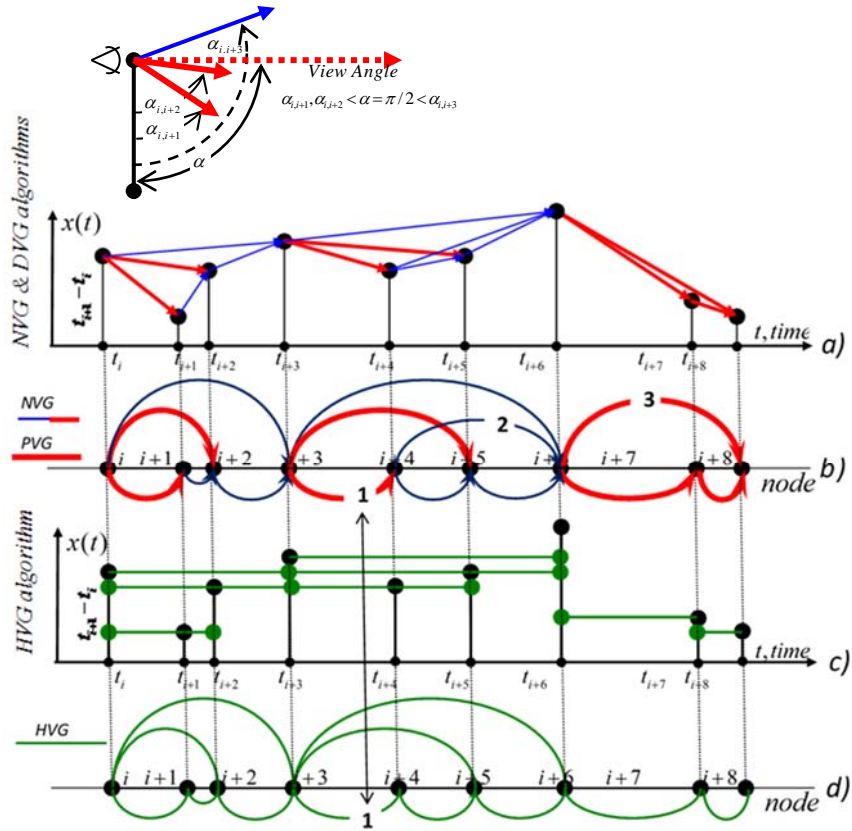

Fig.1. Illustration of the Visibility Graph algorithms. Time series plot for NVG and PNVG algorithms (a) and HVG algorithm (c), corresponding graphs (b) and (d). On the time axis (a) and (c) events $t_i..t_{i+8}$ are marked. Upper left – the PNVG link selection criterion for view angle $\alpha = \pi/2$ applied to node $x(t_i)$. Lines $t_i, t_{i+1}$ and $t_i, t_{i+2}$ with $\alpha_{i,i+1}$ and $\alpha_{i,i+2}$ (thick lines, red online) belongs to the PNVG($\pi/2$), the link with $\alpha_{i,i+3}$ (thin lines, blue online) does not belong to the PNVG($\pi/2$). The NVG (b) consists of thick (red online) and thin (blue online) links. The PNVG($\pi/2$) (b) consists only of thick (red online) links. On the bottom (d) HVG links (green online) are shown. Marked links: 1 – belongs to NVG, HVG and the PNVG($\pi/2$), because its view angle is less than $\pi/2$, 2 - belongs only to the NVG, because its angle is greater than $\pi/2$, 3 – the link that belongs to the NVG and the PNVG($\pi/2$) but not to the HVG.

Note that at the angle of view $\alpha = \pi$ the graphs NVG and PNVG($\pi$) are equal (have the same links). However the PNVG($\pi/2$) does not turn into the HVG. From our point of view, this fact emphasizes the difference between the NVG and the HVG mapping criteria. For example, Fig.1 shows three marked links. Link 1 belongs to all graphs – the NVG, the PNVG($\pi/2$) and the HVG, link 2 belongs to the NVG while this link is not present in the PNVG($\pi/2$) and HVG, the link 3 belongs to the NVG and the PNVG($\pi/2$), but not to the HVG.

Finally, it should be noted that it is possible to compose other parametric mapping algorithms, which take into account other geometrical features of the time series. For example, $x(t_i)$ are built according to the above described procedure but $t_i$ is arithmetic progression with fixed common difference etc.

### III. PNVG PROPERTIES AND ARTIFICIAL DISTRIBUTIONS

Many properties can be considered for complex networks (graphs) [15]. We will discuss here three of them: conventional nodes degree, length of link, which can be defined for any visibility graph [21, 22] and quantity of clusters (connected components of graph), which is specific only to PNVG. To allow the correct comparison between different time series we use relative values with respect to underlying NVG, details can be seen below:

The relative average degree of the nodes - $K(\alpha)$. We denote the average degree of the nodes of the PNVG($\alpha$) as $\overline{k}_{\alpha=\pi}$. In the case of $\alpha = \pi$ i.e. when the PNVG becomes underlying NVG, the property $\overline{k}_{\alpha=\pi}$ becomes a conventional average degree of the nodes of the NVG. The relative average degree of nodes for PNVG($\alpha$) is

$$K(\alpha) = \overline{k}_\alpha / \overline{k}_\pi \tag{4}$$

In the case of $\alpha = \pi$ when the PNVG($\alpha$) turns into the NVG $K(\pi) = 1$ and for low view angles: $K(0) = 0$ and $K(\alpha < \pi/4) = 0$ if all $x(t_i)$ are positive.

The relative average link length - $\Lambda(\alpha)$. This property is specific to any visibility graph [21,22] as well as to PNVG($\alpha$). The link length is the time interval that separates two mutually visible events $i$ and $j$, i.e. if a link $(i, j)$ belongs to the graph then the link length is equal to $t_j - t_i$. The link length can be considered as alternative weight for NVG links also. We denote the average link length of the PNVG($\alpha$) as $\overline{l}_\alpha$. The relative average link length is

$$\Lambda(\alpha) = \overline{l}_\alpha / \overline{l}_\pi \tag{5}$$

Regular values are $\Lambda(\pi)=1$ and $\Lambda(0)=0$, $\Lambda(\alpha<\pi/4)=0$ if all $x(t_i)$ are positive.

The relative number of clusters (connected components of the graph) - $Q(\alpha)$. This property is a specific to PNVG($\alpha$), because NVG has only one connected component. $Q(\alpha)$ is cluster quantity $\overline{q}_\alpha$ divided by number of nodes $N$ in the graph, i.e. $Q(\alpha) = \overline{q}_\alpha / N$

$$Q(\alpha) = \overline{q}_\alpha / N \qquad (6)$$

The PNVG($0$) has no links, i.e. $Q(0)=1$, for positive $x(t_i)$ PNVG($\alpha < \pi/4$) also has no links and $Q(\alpha < \pi/4)=1$, for $\alpha = \pi$ only one cluster is present - $Q(\alpha = \pi) = 1/N$, for other angles of view many clusters exist and $1/N < Q(\alpha) < 1$.

Also we will focus on quantity $Q_{-1}(\alpha)$ - the relative number of clusters with size larger than one node. For any time series PNVG ($0$) have only one node clusters $Q_{-1}(0)=0$, at a maximum angle there exists only one cluster i.e. $Q_{-1}(\pi) = 1/N$. Consequently $Q_{-1}(\alpha)$ always has a maximum. Initially clusters quantity $Q_{-1}(\alpha)$ is growing as more as view angle $\alpha$ is getting wider and more links of the NVG are coming to the PNVG($\alpha$), but at a certain angle $\alpha$ new link connects the disconnected clusters rather than makes cluster size bigger. In other words, at the low view angles the cluster growth process wins, then at the wider angles association process drives and, finally, at the angle of view more than $\alpha \geq \pi/2$ all the clusters join into one.

We used three artificial distributions to investigate the behavior of $K(\alpha)$, $\Lambda(\alpha)$ and $Q(\alpha)$: the uniform random distribution (7a), the Poisson distribution (7b) of time intervals and the Weierstrass distribution (7c) having a fractal dimension $D$. Let $\{r_i, i=1..N+1\}$ be a random value uniformly distributed on the interval $[0..1]$.

$$d_i = r_i, \qquad t_1 = 0,\ t_i = t_{i-1} + d_{i-1}, \qquad x(t_i) = d_i \qquad (7a)$$

$$d_i = -1/\lambda \ln(r_i), \qquad t_1 = 0,\ t_i = t_{i-1} + d_{i-1}, \qquad x(t_i) = d_i \qquad (7b)$$

$$d_i = \sqrt{2}\sigma \frac{\sqrt{1-b^{2D-4}}}{\sqrt{1-b^{(2D-4)(M+1)}}} \sum_{m=0}^{M} \left[ b^{(D-2)m} \sin\left(2\pi(sb^m i + r)\right) \right] \qquad (7c)$$
$$t_1 = 0,\ t_i = t_{i-1} + |d_i|, \qquad x(t_i) = d_i$$

The following parameter values were chosen: for the Poisson distribution: (7b) $\lambda = 1$, for the Weierstrass distribution (7c): $D = 1.3, \sigma = 3.3, b = 2.5, s = 0.005, M = 10$. The Weiesrstrass distribution is alternating, and therefore in (7c) for time we use modulus.

Time series generated according to artificial distributions (7a-7c) plotted in Fig.2.

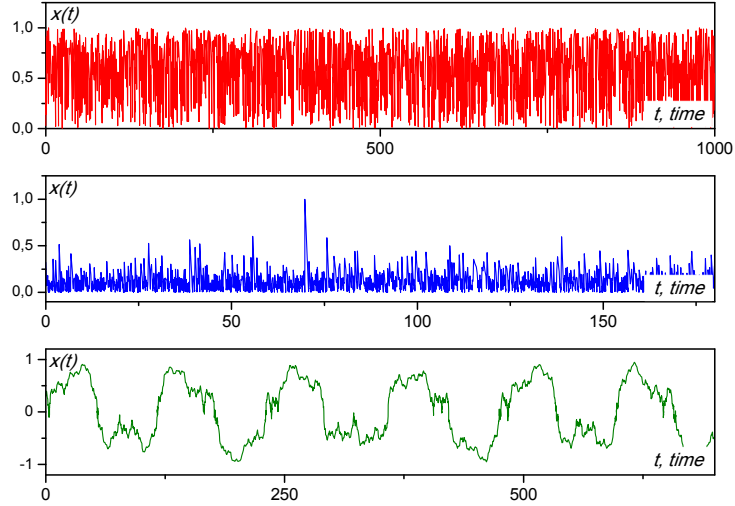

*Fig.2 Time series created by artificial distributions: the random uniform distribution (a), the Poisson distribution (b) and the Weierstrass distribution (c).*

Time series presented in Fig.2 have 2000 points but every distribution generates its own time spread defined by the mean value of time series, uniform distribution (7a) has mean value 0.5 therefore time spread for 2000 points (see Fig.2.a) is close to 1000.

## IV.    RESULTS FOR ARTIFICIAL DISTRIBUTIONS

Calculations were performed in the following way:

First, the time series of length $N = 10^5$ was created using certain distribution (7a-7c). Then above time series was mapped to PNVG($\alpha$) according to rules 1-3 at fixed view angle $\alpha$ in the range $\alpha = [\pi/4, \pi]$. Calculations were performed for 135 equally distributed points within the above range.

Every property of PNVG($\alpha$) - $K(\alpha)$, $\Lambda(\alpha)$ and $Q(\alpha)$ was found by averaging 10 different realizations for every distribution, relative standard deviation (%RSD) for $K(\alpha)$, $\Lambda(\alpha)$ or $Q(\alpha)$ is less than 1% for all data points presented below.

Actually the NVG mapping procedure (rule 1) and weight and direction assignment (rule 2) were performed once for every realization of time series and further it was the base to generate PNVG($\alpha$) at different view angles $\alpha$ (rule 3). The procedure is quite time consuming, its duration depends proportionally upon the length of time series and NVG average node degree $\overline{k}_{\alpha=\pi}$ because all links to be considered when building next PNVG($\alpha$). For random uniform distribution (7a) $\overline{k}_{\alpha=\pi} \approx 5.17$, for Poisson distribution (7b) $\overline{k}_{\alpha=\pi} \approx 5.86$ and the Weierstrass distribution (7c) $\overline{k}_{\alpha=\pi} \approx 22.8$. Calculation time for the Weierstrass distribution is roughly 4 times longer than for uniform and Poisson distributions.

Fig.3 shows the relative average degree of the nodes $K(\alpha)$.

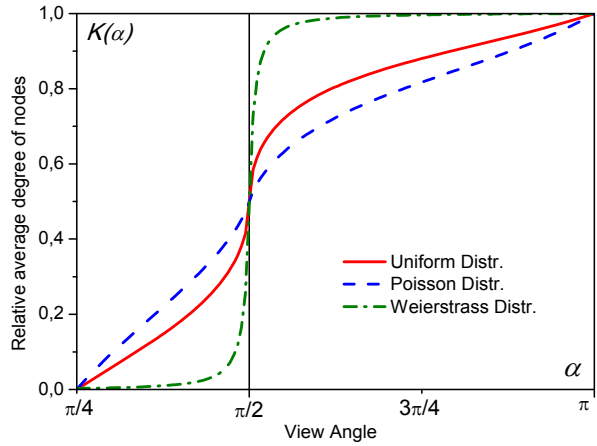

Fig.3 Relative average degree of the nodes $K(\alpha)$ as a function of view angle $\alpha$ for artificial time series: the random uniform distribution (7a), the Poisson distribution (7b) and the Weierstrass distribution (7c).

We denote the distribution of visibility angles of links in PNVG ($\pi$) as $\Gamma(\alpha) = \int_0^\alpha \gamma(x)dx$, where $\gamma(\alpha)$ is corresponding probability density distribution of the angles. The increase of view angle $\alpha$ means more links in PNVG ($\alpha$) and accordingly proportional increase in the average degree of a nodes $\bar{k}_\alpha$ in PNVG($\alpha$). Thus $K(\alpha) = \varepsilon_\Gamma \Gamma(\alpha)$, where $\varepsilon_\Gamma$ is constant. Qualitatively, the angle distribution $\Gamma(\alpha)$ can be represented as a sum of pairwise distribution angles $\Gamma_1(\alpha)$ between adjacent nodes in PNVG ($\pi$) and far-spaced nodes $\Gamma_m(\alpha)$, $m$ - the distance between the nodes $m > 1$.

For uniform time series (7a) adjacent nodes distribution is easy to obtain (see for example [38]) as joint distribution of independent neighbor time intervals

$$\Gamma_n(\alpha) = \begin{cases} -\pi/2 < (\alpha - \pi/2) < 0 & \dfrac{tg(\alpha - \pi/2) + 1}{2}, \\ (\alpha - \pi/2) = 0 & \dfrac{1}{2}, \\ 0 < (\alpha - \pi/2) < \pi/2 & 1 - \dfrac{1}{2(tg(\alpha - \pi/2) + 1)}. \end{cases} \quad (8)$$

For nodes at a distance $m$ from each other (excluding visibility rules 2), using the same method [38], we obtain

$$\Gamma_m(\alpha) = \begin{cases} atcrg(-1/m) < (\alpha - \pi/2) < 0 & (1 + m \cdot tg(\alpha - \pi/2))^2 / 2, \\ \alpha - \pi/2 = 0 & 1/2, \\ 0 < (\alpha - \pi/2) < arctg(1/m) & 1 - (1 - m \cdot tg(\alpha - \pi/2))^2 / 2. \end{cases} \quad (9)$$

With that, distance $m$ more, the smaller is the width of the distribution $[-1/m...1/m]$.

Summing up these distributions we find that $K(\pi/2) = 1/2$ and the slope $K(\alpha)$ is maximal at $\alpha = \pi/2$, $\partial K(\alpha)/\partial \alpha \big|_{\alpha = \pi/2} \to \infty$. Such behavior of $K(\alpha)$ near $\alpha = \pi/2$ will not depend on distribution of the original time series, since the primary role at $\alpha = \pi/2$ will play $\Gamma_m(\alpha)$.

Note also that behavior of $K(\alpha)$ at $\alpha \to \pi$ and $\alpha \to 0$ (for positive $x(t_i)$ $\alpha \to \pi/4$) will be determined by original time series, by distribution $\Gamma_1(\alpha)$. So much for the uniform distribution (7a) from (8) we find that $\partial K(\alpha)/\partial \alpha \big|_{\alpha = \pi/4} > \partial K(\alpha)/\partial \alpha \big|_{\alpha = \pi}$ that is also seen in Fig. 3.

Fig.3 presents the relative link length $\Lambda(\alpha)$ of the angle of view $\alpha$ for distributions (7a-7c).

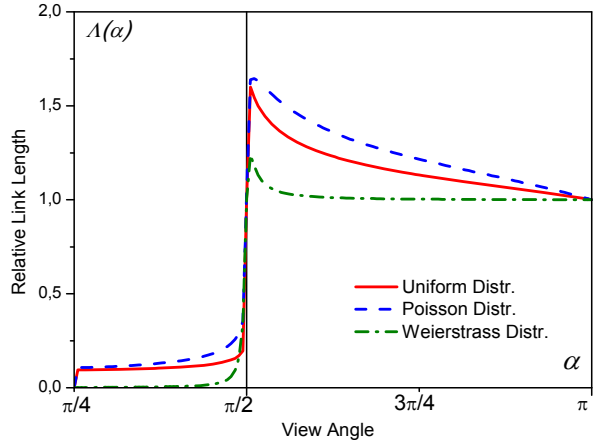

*Fig. 4 Average link length $\Lambda(\alpha)$ $K(\alpha)$ as a function of view angle $\alpha$ for artificial time series: the random uniform distribution (7a), the Poisson distribution (7b) and the Weierstrass distribution (7c).*

As seen from Fig.4 near to $\alpha = \pi/2$ the maximum $\Lambda(\alpha)$ is present. This peak in the angular dependence of $\Lambda(\alpha)$ is due to the fact that for view angles $\alpha \geq \pi/2$ the new long "clear away" links appear like $t_i - t_{i+3}$ (see Fig.1). As soon as the angle becomes wider, many short links are to be added to the PNVG($\alpha$) like the link $t_{i+2} - t_{i+3}$ or $t_{i+4} - t_{i+5}$ (see Fig.1) that lowers calculated average link length $\Lambda(\alpha)$, finally $\Lambda(\pi) = 1$.

This peak (height, shape) can be used to determine (at least partially) the distribution parameters of the original time series.

Fig.5 shows the relative number of the clusters $Q(\alpha)$ in the PNVG($\alpha$) as a function of view angle $\alpha$.

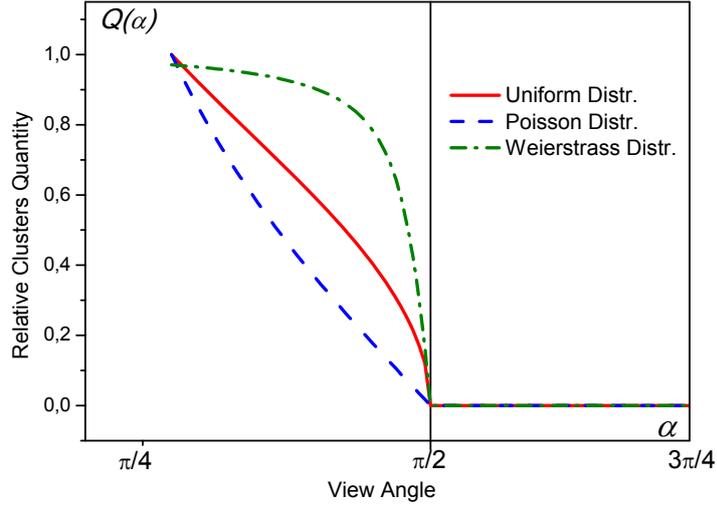

*Fig. 5 Number of clusters $Q(\alpha)$ as a function of view angle $\alpha$ for artificial time series: the random uniform distribution (7a), the Poisson distribution (7b) and the Weierstrass distribution (7c).*

Relative number of clusters $Q(\alpha)$ in PNVG ($\alpha$) at $\alpha \geq \pi/4$ related to average degree of node $K(\alpha)$ as follows:

$$Q(\alpha) \approx 1 - \varepsilon_K K(\alpha), \qquad (10)$$

where $\varepsilon_K$ is the constant.

Indeed, at $\alpha \geq 0$ (for positive $x(t_i)$ $\alpha \geq \pi/4$) when view angle is increasing neighboring single-node clusters are combined to form two-node clusters. The average degree of a node $K(\alpha)$ will grow proportionally to the number of two-node clusters, and therefore the relative number of clusters $Q(\alpha)$ will reduce proportionally.

To see this we have look on Fig.3 that shows the time series with the distributions (7a-7c). Near $\alpha \geq \pi/4$ the dependence $K(\alpha)$ for random uniform distribution (7a) and for Poisson distribution (7b) is close to linear, the corresponding part of the dependence $Q(\alpha)$ in Fig.5 in the area close to $\alpha \geq \pi/4$ also close to linear for the distributions (7a, 7b). In general for distribution (7c) the relation (10) also holds.

On the contrary at the angles $\alpha \approx \pi/2$ PNVG($\alpha$) is formed by one or a few clusters of large volume, at $\alpha > \pi/2$ they are merged into one cluster and the relation similar to (10) does not exist.

Shapes of curves in Fig. 2-5 show that investigated artificial time series (7a-7c) have different behavior of its PNVG($\alpha$) parameters and above can be possible base to determine type of time series distribution and its parameters. Article size does not allow us to present other parameters of PNVG ($\alpha$), which were calculated by the authors. Among these parameters are different relative averages calculated for complex networks, such as connectivity and assortativity of generated graphs, etc.

## V.  RESULTS FOR CARDIAC RHYTHM DATA

Next we will give a short example of the application of the PNVG($\alpha$) algorithm to experimental data – the series of the human heart beat RR intervals. Data was taken from the PhysioNet [39] database. 54 series for healthy people (dataset nsr2db), 25 series for patients with Congestive Heart Failure (chr2db) and 83 series with Atrial Fibrillation (ltafdb) were examined. Each series has different lengths and contains $6 \div 12 \times 10^4$ RR intervals.

ECG data were processed by the PhysioToolkit (*WFDB*) software, provided by PhysioNet [39]. Initially (when necessary) we annotate data using *wqrs* utility, and then to extract series of RR intervals we use *ann2rr* utility. To remove the trend from RR intervals time series we use the detrending algorithm based on smoothness priors approach [40, 41]. The regularization parameter was set to $\lambda_{reg} = 15$. Finally, we correct RR intervals time series so that their mean value is equal to 1.

For each time series above, for view angles $\alpha$ in range $[\pi/4...\pi/2]$ we construct PNVG ($\alpha$) and compute its parameters, particularly $Q_{-1}(\alpha)$. Then we calculate average values over each dataset type. As it was stated before dependence $Q_{-1}(\alpha)$ upon $\alpha$ has maximum. We found positions of such maximum for each time series and its mean value for particular dataset type.

Fig. 6 shows the relative number of the clusters $Q_{-1}(\alpha)$ in the PNVG($\alpha$) as a function of the angle of view $\alpha$.

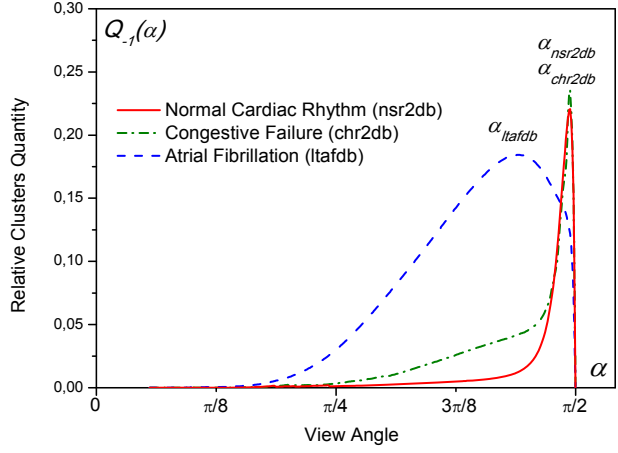

*a*

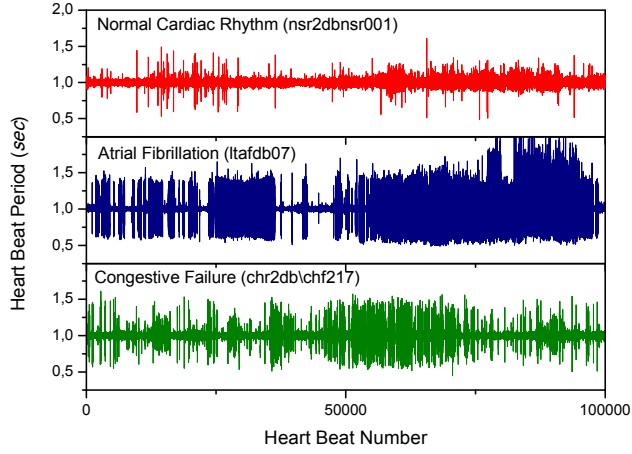

*b*

*Fig. 7. Relative quantity of clusters $Q_{-1}(\alpha)$ as a function of the angle of view $\alpha$ for different RR intervals time series (a), examples of RR intervals time series for different types of cardiac rhythms ( $10^5$ sequential heart beats) (b).*

Each type of cardiac rhythm produces its particular $Q_{-1}(\alpha)$ average curve shape (see Fig. 6) that is visually distinguishable one from another. The shape and position of the maximum of $Q_{-1}(\alpha)$ may be used for identifying the type of cardiac rhythm.

Unfortunately, each RR intervals time series produces its own $Q_{-1}(\alpha)$ shape that is often significantly different from the average shape shown in Fig.6, but there is always at least one peak (sometimes several peaks). For normal cardiac rhythm peak position is $\alpha_{nsr2db} = 0.774 \pm 0.003$, for Congestive Heart Failure - $\alpha_{chr2db} = 0.769 \pm 0.020$ and for Atrial Fibrillation always exists significant peak located at $\alpha < 0.73$ for this peak $\alpha_{ltafdb} = 0.667 \pm 0.023$ (see Fig. 6).

## VI. DISCUSSIONS AND CONCLUSIONS

All the PNVG($\alpha$) properties shown and not shown above, both artificial (Fig.3-5) and experimental (Fig.6) allow us to distinguish, identify and describe various time series. Moreover the parametric visibility algorithm allows to define and evaluate new "parametric" or "dynamic" properties of the built graphs that do not exist in the "static" graph.

It should be noted that the behavior of $Q(\alpha)$ (see Fig.6) near $\alpha = \pi/2$ can demonstrate a power law behavior that could be background to observe similarity to percolation transition or second order phase transition in investigated PNVG($\alpha$). A detailed study of this phenomenon is beyond the scope of this paper, but some data can be found in [42].

In this paper we presented the PNVG algorithm that maps the time series to a dynamic visibility graph using inter events intervals in such a way, that created PNVG($\alpha$) is affine invariant i.e. PNVG($\alpha$) does not depend on the units of original time series. The PNVG algorithm is also able to deal with general type time series like earth quake magnitudes, stock market indices, solar flares etc. Such data needs appropriate modifications of PNVG algorithm to keep universality of view angle parameter.

This article does not cover the full range of possible properties of the PNVG($\alpha$) or all the possible criteria how to select the NVG links for building the PNVG($\alpha$), but it shows the way how to add a new arbitrary varying parameter to the existing non-parametric static algorithm.

Parameters such as view angle $\alpha$ give a new approach to characterize the structure of the time series that are invisible in the conventional version of the algorithm. It is also shown that the PNVG approach allows us to distinguish, identify and describe in detail various time series.